\documentclass[11pt, a4paper, amsfonts, amssymb, amsmath, preprint, showkeys, nofootinbib, twoside]{revtex4-1}
\usepackage[english]{babel}
\usepackage[utf8]{inputenc}
\usepackage[colorinlistoftodos, color=green!40, prependcaption]{todonotes}
\usepackage{amssymb}
\usepackage{amsmath} 
\usepackage{braket}
\usepackage{bbold}
\usepackage{amsthm}
\usepackage{mathtools}
\usepackage{physics}
\usepackage{xcolor}
\usepackage{graphicx}
\usepackage[left=23mm,right=13mm,top=35mm,columnsep=15pt]{geometry} 
\usepackage{adjustbox}
\usepackage{placeins}
\usepackage[T1]{fontenc}
\usepackage{lipsum}
\usepackage{csquotes}
\usepackage[pdftex, pdftitle={Article}, pdfauthor={Author}]{hyperref} 
\usepackage{bm}
\usepackage{appendix}
\usepackage{upgreek}

\begin{document}

\title{Excitons in twisted AA' hexagonal boron nitride bilayers}

\author{Pedro Roman-Taboada, Estefania Obregon-Castillo, Andrés R. Botello-Mendez, Cecilia Noguez}

\email[Email:]{cecilia@fisica.unam.mx}
\affiliation{Instituto de Física, Universidad Nacional Autónoma de México, Apartado Postal 20-364, Ciudad de Mexico 01000, Mexico}

\date{\today} 

\begin{abstract}
The twisted hexagonal boron nitride (hBN) bilayer has demonstrated exceptional properties, particularly the existence of electronic flat bands without needing a magic angle, suggesting strong excitonic effects. Therefore, a systematic approach is presented to study the excitonic properties of twisted AA’ hBN using the Bethe-Salpeter equation based on single-particle tight-binding wave functions. These are provided by a one-particle Hamiltonian that is parameterized to describe the main features of {\it ab initio} calculations. The Bethe-Salpeter equation is then solved in the so-called excitonic transition representation, which significantly reduces the problem dimensionality by exploiting the system’s symmetries. Consequently, the excitonic energies and the excitonic wave functions are obtained from the direct diagonalization of the effective two-particle Hamiltonian of the Bethe-Salpeter equation. We have studied rotation angles as low as $7.34^{\circ}$. The model allows the study of commensurate and incommensurate moiré patterns at much lower computational cost than the {\it ab initio} version of the Bethe-Salpeter equation.  Here, using the model and effective screening of the Keldysh type, we could obtain the absorption spectra and characterize the excitonic properties of twisted hBN bilayers for different rotation angles, demonstrating how this property affects the excitonic energies and localizations of their wavefunctions.
\end{abstract}


\maketitle
\section{Introduction}
Two-dimensional (2D) materials and their heterostructures are ideal candidates for many technological applications \cite{novoselov20162d,lee2016reconfigurable,chaves2020bandgap,liu2021promises}. In particular, they are suitable for nanophotonics applications \cite{low2017polaritons,autere2018nonlinear,guo20192d}. These kinds of materials can give rise to strong light-matter interactions through a myriad of dipole-type polaritonic excitations, such as infrared-active phonons \cite{kuhlmann1998infrared}, excitons in 2D semiconductors \cite{ross2013electrical,xiao2017excitons,mak2018light}, and plasmons in doped 2D materials \cite{wang2020optical}. In 2D semiconductors, the appearance of strongly bound excitons opens the possibility of realizing efficient energy transfer by driving charge excitons via applied electric fields \cite{liu2017electric,cavalcante2018stark}. Furthermore, they are used in solar cells \cite{song2013external} and photodetectors \cite{riis2018efficient}. In this context, a 2D hexagonal boron nitride (hBN) monolayer is a semiconductor with a wide band gap, which exhibits strong correlation effects. For instance, excitons are characterized by binding energies of about $2$ eV, which makes hBN an excellent candidate for applications in optoelectronic devices in the deep-ultraviolet region \cite{dahal2011epitaxially,li2012dielectric}. On the other hand, heterostructures made out of 2D hBN monolayers are also of great interest. For example, by twisting an hBN bilayer, one can obtain flat bands without the necessity of magic angles \cite{zhao2020formation,xian2019multiflat} due to an enhancement of the correlation effects, thus, improving its excitonic properties. 

The Bethe-Salpeter equation is the most common approach to study the excitonic phenomena in condensed matter systems. It is a four-point equation that allows the calculation of the two-particle correlation function and describes the propagation of two particles within the Green's function formalism \cite{martin2016interacting}. Its $\bm{k}$-space representation is often used to deal with periodic systems \cite{leng2016gw,rohlfing1998electron}. Unfortunately, the {\it ab initio} form of the Bethe-Salpeter equation can be extremely cumbersome and hard to solve even for periodic systems with few atoms per unit cell. For example, commensurate twisted hBN bilayer, whose excitonic properties are impossible to compute by this means even for its smallest unit cell. It translates into the need for sophisticated numerical approaches and the necessity of huge computational resources \cite{schmidt2003efficient,marini2009yambo,blase2011first,vorwerk2019bethe}.

Typically, one first computes the quasiparticle band structures fed into the Bethe-Salpeter equation using a combination of Density Functional Theory (DFT) and many-body perturbation theory \cite{vinson2011bethe,blase2020bethe,henriques2022excitonic} ({\it e.g.}, GW approximation) and time-dependent DFT \cite{ramasubramaniam2019transferable,suzuki2020excitons,camarasa2023transferable}. Instead of performing these expensive band structure calculations, one can directly employ a continuum or a tight-binding model to get them, significantly reducing the complexity of the problem. The continuum model works well for twisted systems with extensive unit cells or, equivalently, minimal rotation angles \cite{tarnopolsky2019origin,devakul2021magic}. In this case, an effective continuum Hamiltonian can be written in $\bm{k}$-space that is often expanded in Bloch states, forming a momentum lattice that is then truncated \cite{carr2020electronic}. This approach allows fast and precise calculation of the band structure of the twisted systems within a reduced range of energies. However, it is still being determined if the excitonic properties are correctly described since the model needs to include a detailed local atomic description. This latter is crucial for the correct computation of the dielectric function. Despite this, the approach has shown to be a good approximation for the case of non-twisted AA' and AB hBN bilayers but might fail to predict their optical response satisfactorily \cite{henriques2022excitonic}.

Using single-particle tight-binding states as input for the Bethe-Salpeter equation in real space has proven very successful \cite{delerue2000excitonic,jiang2007chirality,galvani2016excitons,sponza2018direct,paleari2018excitons}. Here, we employ a parameterized single-particle tight-binding model to include quasiparticle corrections. When this approach is applied to periodic systems, their symmetries are exploited to reduce the problem’s dimensionality remarkably. This latter is possible because the number of electron-hole pairs needed to solve the Bethe-Salpeter equation can be hugely reduced by considering only non-equivalent hole sites \cite{galvani2016excitons}. Then, one can use the excitonic transition basis, built on single-particle wave functions, to write down the effective two-particle Hamiltonian of the Bethe-Salpeter equation.

Taking all this into account, the aim of this paper is twofold. First, we present a systematic approach to studying the excitonic properties of twisted AA' hBN bilayers using the Bethe-Salpeter equation based on adequate tight-binding models. Using moderate computational resources, the model allows us to analyze systems with unit cells with thousands of atoms. For the sake of simplicity, we describe the Bethe-Salpeter kernel, {\it i.e.}, the direct Coulomb interaction and the exchange effects, by a model potential or, in other words, an effective screening. This approximation is only valid when the model potential is relatively smooth. To solve the Bethe-Salpeter equation, we write it down into the basis of excitonic transitions, which are constructed using the tight-binding single-particle wave functions. Second, we apply the model to twisted AA' hBN bilayers for several rotation angles to study their excitonic properties. As a remark, even though we have employed our model to hBN bilayers of the AA' type, its application to other vertical stacking systems is straightforward. Also, considering other 2D crystalline structures is possible as long as suitable tight-binding models and a smooth electron-hole interaction model potential are available.

This work  is organized as follows. Section \ref{single-partice_model} describes the single-particle model that defines the electronic properties of non-twisted and twisted AA' hBN bilayers. Section \ref{Bethe-Salpeter_equation} is devoted to presenting the details of the two-particle model. All the approximations used here are listed, and the explicit form of the real space representation of the Bethe-Salpeter equation is shown. Section \ref{excitonic_wf} presents the results obtained for commensurate twisted AA' hBN bilayers, and the discussion. The conclusions are set down in section \ref{sec:conclusions}. Finally, most of the technical information is given in the appendix.

\section{Single-particle model}
\label{single-partice_model}
We define the tight-binding Hamiltonian used to describe the one-particle electronic properties of a non-twisted AA' hBN bilayer. To do so, we will assume that the electronic properties of an hBN bilayer at low energies are well described, within the one-particle approximation, by a $p_z$ tight-binding Hamiltonian, $H^{\text{1P}}$. In the second quantization formalism, $H^{\text{1P}}$ takes the following form,
\begin{equation}
\begin{split}
    H^{\text{1P}} =  &-t_{\parallel}\sum_{\left< i,j\right >,\alpha}\left( c^{\dag}_{\alpha,i}c_{\alpha,j}+ \text{h.c.}\right ) \\ & - t_{\perp}\sum_{i,j}^M\left (c^{\dag}_{1,i}c_{2,j} + \text{h.c.}\right ) \\
& + \sum_{i}^{M/2}\left[\Delta_B \left(c_i^B\right)^{\dag}c^{B}_{i} +\Delta_N \left(c^{N}_{i}\right)^{\dag}c^{N}_{i}\right],
\end{split}
\label{h1p}
\end{equation}
where $\left< i,j\right >$ indicates summation over nearest and next-nearest neighbors, $c_{\alpha,i}$ is the annihilation operator for a $p_z$ electron at the site $i$ within the $\alpha$ layer, with $\alpha = 1, 2$. The operator $c_i^B$ ($c_i^N$) annihilates an electron at the $i$-th boron (nitrogen) site. Therefore, $\Delta_N$ and $\Delta_B$ are the  on-site energies of nitrogen and boron atoms, respectively. Finally, $M$ is the number of atoms in the system, and $t_{\parallel}$ is the intralayer hopping parameter, {\it i.e.}, the interaction between atoms within the same layer. At the same time,  $t_{\perp}$ stands for the interlayer hopping parameter, {\it i.e.}, the interaction between atoms at different layers.

Here, intralayer hopping parameters are considered constant, although interactions up to second-nearest neighbors are included. As a result, we have three types of interactions, namely, boron-nitrogen ($t_{\parallel}^{\text{BN}}$), nitrogen-nitrogen ($t_{\parallel}^{\text{NN}}$), and boron-boron ($t_{\parallel}^{\text{BB}}$) interactions. In contrast, to correctly reproduce the electronic properties of hBN bilayers, the interlayer hopping parameters must follow $V_{pp}$ interactions,
\begin{equation}
    \begin{split}
        t_{\perp}(r) & = n^2 V_{pp\sigma} e^{-\beta_1(r - c_0)/c_0} \\
        & + (1 - n^2)V_{pp\pi}e^{-\beta_2(r-c_0)/c_0},
    \end{split}
\end{equation}
where $n = \bm{r}\cdot\bm{e}_z/r$ is the direction cosine of the vector $\bm{r}$ that joins two atoms at different layers and $r = \left|\bm{r}\right|$. The parameters $V_{pp\pi}$, $V_{pp\sigma}$, $\beta_1$, and $\beta_2$ are fit to {\it ab initio} calculations. As for intralayer hopping parameters, we have three types of interlayer hopping parameters: $t_{\perp}^{\text{BN}}$, $t_{\perp}^{\text{NN}}$, and $t_{\perp}^{\text{BB}}$. Additionally, all interactions are zero for distances bigger than a cutoff radio, $r_{\text{cutoff}} = 7$ \AA. The best-fit values of all these parameters are listed in Table \ref{oneparticle_parameters}.  In the appendix, a detailed explanation of how the parameters were tested for different hBN bilayers and the errors associated with the parameters, listed in Table \ref{table_par}, can be found.

\begin{table}
\begin{center}
\begin{tabular}{ | c | c | c | c |  }
 \hline
 & BN & NN & BB \\
 \hline
$V_{pp\pi} $ [eV]   & -1.932 & -1.162 &   -1.323\\
 \hline
$V_{pp\sigma}$ [eV] & 0.398  & 0.153 & 0.817 \\
 \hline
$\beta_1$ & 3.739 & 11.91 & 4.156 \\
 \hline
$\beta_2$ & 6.292 & 8.317 & 5.940 \\
\hline
$t_{\parallel}$ [eV] & -2.725 & 0.222 & 0.019 \\
\hline
\end{tabular}
\caption{Parameters used here to describe the one-particle electronic structure of AA' hBN bilayers.}
\label{oneparticle_parameters}
\end{center}
 \end{table}

\subsection{Non-twisted AA' hBN bilayer}

\begin{figure*}
\includegraphics[width=3.25in]{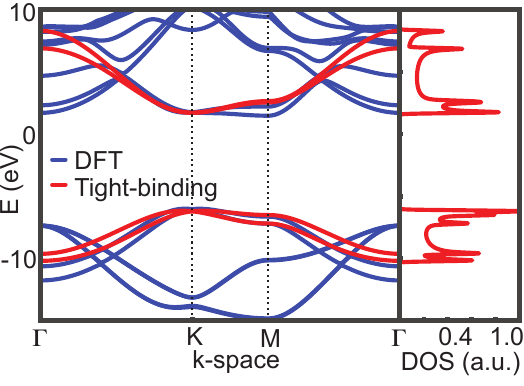}
\caption{Right panel. Single-particle tight-binding (red lines) and DFT (blue lines) electronic band structure of a non-twisted hBN bilayer of type AA'. The tight-biding model reproduces the indirect band gap around the $\textbf{K}$-point. Although this model is not in quantitative agreement with DFT calculations, it qualitatively reproduces their features at low energies, {\it i.e.}, near the $\textbf{K}$ point. Left panel. The corresponding normalized DOS for the tight-binding model.}
\label{1p_bands_zero}
\end{figure*}

To validate the model, we computed the band structure and the density of states (DOS) of a non-twisted AA' hBN bilayer. It was done by directly diagonalizing the periodic version of equation (\ref{h1p}). For this purpose, we used a linear combination of localized atomic orbitals, chosen as Bloch functions. The band structure for a $(\bm{\Gamma}, \textbf{K}, \textbf{M}, \bm{\Gamma})$ $\bm{k}$-path and the corresponding DOS are shown in Fig. \ref{1p_bands_zero}. It can be seen that the tight-binding model (red lines) has an indirect band gap going from some point near the $\textbf{K}$-point at the valence band to the $\textbf{K}$-point at the conduction band, whose size is approximately $7.87$ eV. This latter value was obtained after adjusting the parameters to reproduce the band gap at the $\bm{\Gamma}$-point provided by GW calculations \cite{sponza2018direct,paleari2018excitons}. Besides that, this model qualitatively reproduces the main features found for AA' hBN bilayer in DFT calculations (blue lines in Fig. \ref{1p_bands_zero}, near the $\textbf{K}$-point), as long as a scissor operation is performed to reproduce the GW band gap correctly. The corresponding normalized DOS for our tight-binding model is presented in the right panel of Fig. \ref{1p_bands_zero}. Once the model has been validated, we study the case of twisted AA' hBN bilayers.

\subsection{Twisted AA' hBN bilayer}

\begin{figure}[h]
\includegraphics[width=3.25in]{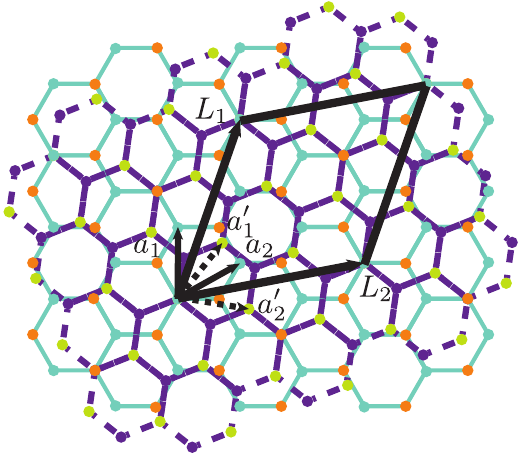}
\caption{The lattice structure of twisted hBN bilayer for $(m_1,m_2) = (2,1)$. Solid blue and dotted violet lines represent the first and second layers, respectively. Here, the second layer is rotated with respect to the first one. The unit vector of layer 1 (layer 2) are denoted by $\bm{a}_1$ and $\bf{a}_2$ ($\bm{a}_1'$ and $\bm{a}_2'$), while the unit vectors of the superlattice ($\bm{L}_1$ and $L_2$) are given by $\bm{L}_1 = m_1\bm{a}_1+m_2\bm{a}_2 $ and $\bm{L}_2=-m_2\bm{a}_1 + (m_1+ m_2)\bm{a}_2$.}
\label{tw_hbn_21}
\end{figure}

Each layer of the bilayer system has a unit cell made of two atoms, one boron and one nitrogen, defined with the following unit vectors,
\begin{equation}
\begin{split}
    \bm{a_1} & = a(1,0) \\
    \bm{a_2} & = a\left(\sqrt{3},1\right)/2.
\end{split}
\end{equation}
The superlattice of a commensurate twisted hBN bilayer can be determined by its unit vectors, which are established via two integers, $m_1$ and $m_2$, as usual \cite{moon2013optical,aragon2019twisted},
\begin{equation}
    \begin{split}
        \bm{L_1} & = m_1\bm{a}_1+m_2\bm{a}_2 \\
        \bm{L_2} & = -m_2\bm{a}_1 + (m_1+ m_2)\bm{a}_2.
    \end{split}
\end{equation}
For example, the twisted AA' hBN bilayer with $(m_1,m_2) = (2,1)$ is shown in Fig. \ref{tw_hbn_21}. The rotation angle between layers can be obtained as follows,
\begin{equation}
    \cos{(\theta)} = \frac{m_1^2 + 4m_1m_2 + m_2^2}{2(m_1^2 + _1m_2 + m_2^2)}.
\end{equation}
In this work, we have studied four pairs of indices $m_1$ and $m_2$, namely, $(2,1)$, $(3,2)$, $(4,3)$, and $(5,4)$. These indices are equivalent to rotation angles given by $\theta = 21.79^{\circ}$, $13.17^{\circ}$, $9.43^{\circ}$, and $7.34^{\circ}$. Although all the calculations were performed using periodic boundary conditions from here on, it is essential to mention that our model can also be applied to incommensurate twisted hBN bilayers.

\begin{figure*}
\includegraphics[width=3.25in]{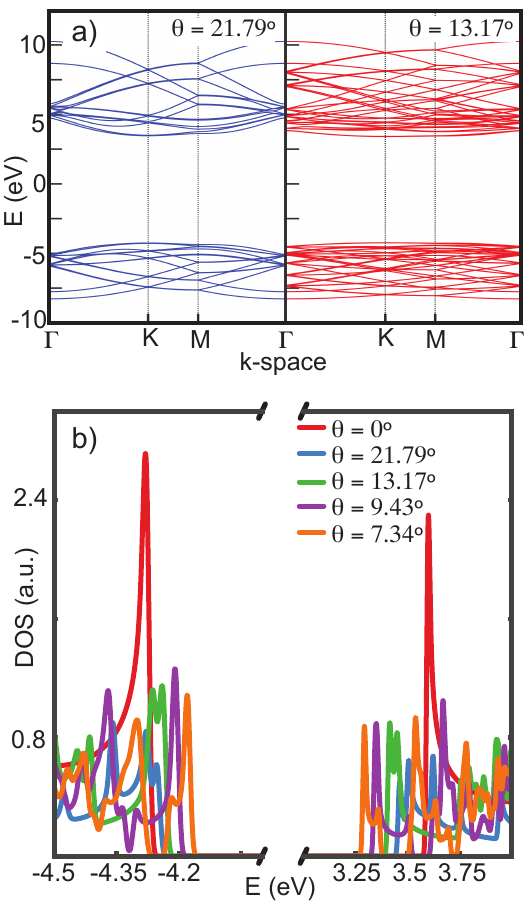}
\caption{Panel a). Single-particle band structures of a twisted AA' hBN bilayer with $(m_1,m_2) = (2,1)$ for the left panel and $(m_1,m_2) = (3,2)$ for the right one. Such band structures were obtained from our one-particle tight-binding model. In the left panel, note how the valence and conduction bands begin to flatten near the $\textbf{K}$-point, being more pronounced on the right panel, where truly flat bands appear. This effect agrees with {\it ab initio} calculations previously reported \cite{xian2019multiflat}. Panel b). In this plot, we display a zoom of the single-particle DOS near the conduction and valence band edge for several rotation angles. In this way, it can be observed how sharp peaks emerge inside the band gap of a non-twisted AA' hBN bilayer as the rotation angle decreases.}
\label{1p_bands_one}
\end{figure*}

The {\it ab initio} calculation shows that flat bands can be obtained without a magic angle by twisting AA' hBN bilayers \cite{xian2019multiflat}. Flat bands emerge at large rotation angles, for example, at $13.17^{\circ}$. Our simple model qualitatively reproduces this feature of twisted AA' hBN bilayers. In Fig. \ref{1p_bands_one} (a), we show the band structure of a twisted hBN bilayer for two pairs of $(m_1,m_2)$ indices, namely $(m_1,m_2) = (2,1)$ and $(3,2)$. Although the band structure for the $(m_1,m_2) = (2,1)$ case does not show any flat band (see left panel of Fig. \ref{1p_bands_one} (a)), note how the states at the $\textbf{K}$-point begin to flatten. In fact, by decreasing the rotation angle to $\theta = 13.17^{\circ}$, real flat bands emerge, as seen in the right panel of Fig. \ref{1p_bands_one}(a), since they are isolated from the other bands by about $0.01$ eV. For rotation angles ($\theta$) smaller than $13.17^{\circ}$, the width of the flat bands diminishes. This feature can be better appreciated by plotting the DOS for various angles, as is done in Fig. \ref{1p_bands_one}(b). Note how sharp peaks emerge within the band gap of a typical AA' hBN bilayer as the rotation angle decreases (see Fig. \ref{1p_bands_one} (b)). For angles smaller than $21.79^{\circ}$, these peaks are a signature of flat bands near the conduction and valence band edges. 

After having presented and proven the validity of our model, we proceed to its application to study the excitonic properties of twisted AA' hBN bilayer. This study is done in the next section, where we briefly review the Bethe-Salpeter equation and introduce our approach to treating excitonic effects on periodic 2D systems.

\section{Bethe-Salpeter equation}
\label{Bethe-Salpeter_equation}
\subsection{Reciprocal space representation}

In condensed matter, the Bethe-Salpeter equation (BSE) describes bound states in a two-body particle system that, in the $\bm{k}$-space, is often written as an effective eigenvalue problem for electron-hole pairs \cite{galvani2016excitons,sponza2018direct},
\begin{equation}
\begin{split}
    & \left(E_{\bm{k}c}^{\text{1P}} - E_{\bm{k}v}^{\text{1P}}\right) \Psi_{\bm{k}vc}^{\lambda} \\ 
    & + \sum_{\bm{k}'v'c'}\left < \bm{k}vc \right | K_{\text{eh}} \left | \bm{k}'v'c' \right > \Psi_{\bm{k}'v'c'}^{\lambda} = E^{\lambda}\Psi_{\bm{k}vc}^{\lambda},
\end{split}
\label{k_bse}
\end{equation}
where $E_{\bm{k}c}^{\text{1P}}$ and $E_{\bm{k}v}^{\text{1P}}$ are the single-particle conduction and valence band energies, respectively. The electron-hole interaction kernel is denoted by $K_{\text{eh}}$, and
\begin{equation}
    \left |\bm{k}n_1n_2\right > = \left |\bm{k}n_1\right>\otimes\left |\bm{k}n_2\right>
\end{equation}
with $\left |\bm{k}n_1\right>$ being a single-particle Bloch function obtained from the one-particle Hamiltonian (\ref{h1p}). Additionally, $E^{\lambda}$ stands for the $\lambda$-th excitonic energy and $\Psi_{\bm{k}vc}^{\lambda}$ are the expansion coefficients of the $\lambda$-th excitonic state $\left | \Psi^{\lambda} \right >$ in terms of electron-hole excitations in the reciprocal space, given by $ \left | \Psi ^{\lambda}\right > = \sum_{vc}\bm{k}\Psi_{\bm{k}vc}^{\lambda}\left | \bm{k}vc \right > $.
Here
\begin{equation}
    \left | \bm{k}vc \right > = a^{\dag}_{c\bm{k}}a_{v\bm{k}}\left | \emptyset \right >,
\label{kvc}
\end{equation}
with $\left | \emptyset \right >$ denoting the vacuum state. For simplicity, the spin of the electrons and holes is not considered, \emph{i. e.}, only singlet states were studied.  This assumption can be justified since triplet states are dark when the spin-orbit coupling is not strong enough to induce spin-flips, as is the case for ideal hBN systems \cite{wirtz2005excitonic}. Additionally, only vertical transitions for the electron-hole pairs have been considered, leading to a vanishing wave vector of its center of mass, $\bm{Q} = 0$ in $\bm{k}_e = \bm{k}_h +\bm{Q} = \bm{k}$. Under these conditions, the $\lambda$-th six-dimensional excitonic wave function can be expressed in terms of Bloch functions and excitonic weights,
\begin{equation}
    \Psi^{\lambda}(\bm{r}_e,\bm{r}_h) = \sum_{vc\bm{k}} \Psi_{\bm{k}vc}^{\lambda} \varphi_{c\bm{k}}(\bm{r}_e)\varphi^*_{v\bm{k}}(\bm{r}_h),
\end{equation}
where $\bm{r}_e$ and $\bm{r}_h$ are the positions of the electron and hole, respectively.

\subsection{Real space representation}

The conduction (valence) states near the band gap are concentrated in boron (nitrogen) atoms for AA' hBN bilayers \cite{galvani2016excitons,xian2019multiflat}. Under these circumstances, one can approximate equation (\ref{kvc}) as,
\begin{equation}
\begin{split}
    \left | \bm{k}vc \right > & \approx a^{\dag}_{B\bm{k}}a_{N\bm{k}}\left | \emptyset \right > \\ 
    & \approx \frac{1}{M}\sum_{\bm{\beta},\bm{\alpha}}a^{\dag}_{B\bm{\alpha}}a_{N\bm{\beta}}e^{i\bm{k}\cdot (\bm{\alpha}-\bm{\beta})}\left | \emptyset \right >.
\end{split}
\label{app_kexc}
\end{equation}
Here, the sum runs over electron ($\bm{\alpha}$) and hole ($\bm{\beta}$) positions, while $N$ and $B$ are nitrogen and boron atoms, respectively. This latter sum can be further simplified by noting that for electron-hole pairs, it is enough to know their relative position, namely, $\bm{R} = \bm{\alpha} - \bm{\beta}$. Hence, the equation (\ref{app_kexc}) sum can be rewritten as follows,
\begin{equation}
    \begin{split}
        \left | \bm{k}vc \right > & \approx \frac{1}{M}\sum_{\bm{\beta},\bm{R}}a^{\dag}_{N+\bm{R},\bm{\beta}}a_{N\bm{\beta}}e^{i\bm{k}\cdot \bm{R}}\left | \emptyset \right > \\
        & \equiv \frac{1}{\sqrt{M}}\sum_{\bm{R}}e^{i\bm{k}\cdot\bm{R}}\left | \bm{R}vc \right >,
    \end{split}
    \label{realbasis}
\end{equation}
where we have defined $\left | \bm{R}vc \right >$ as
\begin{equation}
    \left | \bm{R}vc \right > = \frac{1}{\sqrt{M}}\sum_{\bm{\beta}}a^{\dag}_{N+\bm{R},\bm{\beta}}a_{N\bm{\beta}}\left | \emptyset \right >.
\end{equation}
The states $\left | \bm{R}vc \right >$ are the elementary excitonic states in real space. Equivalently, $\left | \bm{R}vc \right >$ can be seen as the Bloch state describing the motion of an electron-hole pair of size $\bm{R}$.

Let us say a few words about the advantages of this type of basis. First, since the system under study (AA' hBN bilayers) is periodic, it is unnecessary to consider all the electron-hole pairs. By symmetry properties, examining only non-equivalent hole positions within the bilayer is enough, significantly reducing the problem's dimensionality. Second, by restricting electrons to be only at boron sites, the number of electron-hole pairs is decreased again.  This assumption is based on previous results reported in Ref. \cite{galvani2016excitons}, where  authors found that near the gap, the valence (conduction) Bloch states are concentrated on nitrogen (boron) atoms, which has been corroborated in Refs. \cite{paleari2018excitons,sponza2018direct}.

Finally, considering up to first-order effects and neglecting exchange ones, the electron-hole interaction kernel of the BSE is diagonal. Although these approximations seem very drastic, their application to hBN bilayers results in good agreement with more complex and cumbersome models, such as Bethe-Salpeter calculations based on the GW approximation.

We have closely followed the work presented in reference \cite{galvani2016excitons}. However, we improve this model to adequately describe the excitonic properties of twisted AA' hBN bilayers. First, we allow the electrons to be at boron sites and nitrogen ones. By doing this, we expect to correctly consider the complex atomic environment which characterizes twisted bilayer systems. Second, in Ref. \cite{galvani2016excitons}, the kinetic part of the real-space BSE is obtained by first-order perturbation theory, which leads to a next-nearest neighbor approximation in the excitonic transition space. In this work, we follow a different approach, detailed below, that allows us to consider all types of interactions at once, albeit at a higher computational cost. Finally, instead of expanding the BSE Hamiltonian with a set of one-particle tight-binding $s$-orbitals, we use a set of $p_z$ orbitals, which has been demonstrated to reproduce better the results of DFT calculations for non-twisted and twisted AA' hBN bilayers.

We now obtain the explicit representation of the BSE in real space. We substitute equation (\ref{realbasis}) into equation (\ref{k_bse}) to do so. After some manipulations, we obtain,
\begin{equation}
    \begin{split}
        E^{\lambda}\Psi_{\bm{R}vc}^{\lambda} & = \sum_{\bm{R}'}\left < \bm{R}vc\right | H^{\text{1P}}\left | \bm{R}'v'c'\right >\Psi_{\bm{R}'v'c'}^{\lambda} \\
        & + \left < \bm{R}vc \right | K_{\text{eh}} \left | \bm{R}vc \right > \Psi_{\bm{R}vc}^{\lambda}.
    \end{split}
    \label{real_bse}
\end{equation}
To consider all possible interactions included in the kinetic energy term of the previous equation, we compute it as follows,
\begin{equation}
    \begin{split}
        & \left < \bm{R}\right | H^{\text{1P}} \left | \bm{R'} \right > = \sum_{vc}(E_c - E_v) \left < \bm{R} |  vc\right >\left < vc | \bm{R'}\right >.
    \end{split}
    \label{kineticpart}
\end{equation}
We have omitted the $vc$ label in the excitonic states $\left | \bm{R}vc \right > \equiv \left | \bm{R} \right >$ for simplicity. On the other hand,  $\left | vc\right > = \left | v\right >\otimes\left | c\right >$, where $\left | n\right >$ with $n = c,\,v$, is an eigenstate of $H^{\text{1P}}$. In deriving equation (\ref{kineticpart}), we have assumed that the $\left| vc\right>$ states are complete and that the $H^{\text{1P}}$ can be split into two subspaces as $H^{\text{1P}} \approx H_c\otimes\bm{1}_v - \bm{1}_c\otimes H_v$, which is reasonable for 2D semiconductors with a wide bandgap \cite{galvani2016excitons}.

For the electron-hole interaction kernel, we neglect the exchange contributions. In this approximation and the basis $\left |\bm{R}\right >$, the kernel, up to first-order, becomes diagonal. By following reference \cite{galvani2016excitons}, we also assume the kernel can be substituted by a model potential, particularly a potential of the 2D Keldysh type \cite{cudazzo2011dielectric}:
\begin{equation}
    V(r) = \frac{e}{4\pi\epsilon_0r_0}\frac{\pi}{2}\left [ H_0\left(\frac{r}{r_0}\right) - Y_0\left(\frac{r}{r_0}\right) \right ],
\end{equation}
where $r_0$ is the screening length, and $H_0$ and $Y_0$ are the Struve function and the Bessel function of the second kind, respectively. As a reminder, this type of interaction behaves as a screened $1/r$ Coulomb potential at long range. At the same time, it has a weak logarithmic divergence at short distances, where the screening distance, $r_0$, determine the crossover. Thus the matrix elements of the Bethe-Salpeter kernel, in the excitonic transition basis, can be written as,
\begin{equation}
    \left < \bm{R} \right | K_{\text{eh}} \left | \bm{R} \right > = V(\bm{R}).
\end{equation}
Since we are dealing with a bilayer system, two screening lengths are defined, namely $r_0^{\text{IP}}$ and $r_0^{\text{IL}}$. Here, $r_0^{\text{IP}}$ corresponds to the screening length of the Keldysh potential for interactions in a single layer, while $r_0^{\text{IL}}$ stands for the screening length for interactions among different layers. These parameters were adjusted to reproduce the excitonic energies of a non-twisted  AA' hBN bilayer correctly. The best fitting values we have found are $r_0^{\text{IP}} = 0.87$ nm and $r_0^{\text{IL}} = 0.91$ nm.

In the excitonic transition basis, the $\lambda$-th excitonic wave function can be readily written as,
\begin{equation}
    \begin{split}
        \Psi^{\lambda}(\bm{r}_e,\bm{r}_h) = \sum_{\bm{R}}\sum_{\bm{\beta}} \Psi_{\bm{R}}^{\lambda} & \,\varphi_{e}(\bm{r}_e-\bm{\beta}-\bm{R})\times \\
        & \varphi^*_{h}(\bm{r}_h - \bm{\beta}),
    \end{split}
    \label{realexc_wf}
\end{equation}
where $\bm{r}_h$ is the position of the hole, $\bm{r}_e$ is the position of the electron, $\bm{\beta}$ runs over the position of the non-equivalent holes (nitrogen atoms), and $\Psi_{\bm{R}}$ are the excitonic weights obtained from the eigenvectors of the effective two-particle Hamiltonian in equation (\ref{real_bse}).

This work mainly aims to apply the BSE (\ref{real_bse}) to twisted AA' hBN bilayers. For that, we feed the BSE (\ref{real_bse}) with the eigenvalues and eigenvectors of the one-particle Hamiltonian $H^{\text{1P}}$, defined in equation (\ref{h1p}). Before applying our model to twisted systems, we have verified that our model correctly reproduces the results from other authors \cite{galvani2016excitons,paleari2018excitons,sponza2018direct,henriques2022excitonic}.

The path chosen to obtain the excitonic properties of AA' hBN bilayers is stepwise. The first step is building the periodic version of equation (\ref{h1p}) for a given pair of $m_1$ and $m_2$ indices. Then, in the second step, we obtain the matrix representation of the real-space version of the BSE (\ref{real_bse}), taking into account as many holes as non-equivalent positions they are. Remember that we suppose that holes can only live in nitrogen atoms. For example, a non-twisted AA' hBN bilayer will have only two non-equivalent holes or nitrogen atoms, one per layer. However, the system's symmetry decreases as the rotation angle decreases. In other words, the unit cell becomes larger, giving rise to more non-equivalent sites for nitrogen atoms.

At this point, how the effective eigenvalue problem scales upon the considered number of holes must be clarified. For simplicity, let us consider the case of a non-twisted hBN bilayer. As we have mentioned before, the number of non-equivalent holes ($n_{\text{h}}$) for this system is two. Since the electron can be at any atomic site within the bilayer, we have $n_h M$ electron-hole pairs, where $M$ is the total number of atoms in the bilayer. Therefore, the matrix representation of equation (\ref{real_bse}) has dimension $n_{\text{h}}M\times n_{\text{h}}M$. However, due to the sum over valence and conduction states appearing in the kinetic part of the equation (\ref{real_bse}), we need to sum $(M/2)^2$ items per matrix element. In this way, the building of the Bethe-Salpeter effective two-particle Hamiltonian scales as $n_\text{h}^2 M^4/4$. As the prefactor $n_\text{h}$ grows, more computational resources are needed. For the rotation angles studied here, the number of non-equivalent holes is $n_h = 3$, $7$, $13$, and $21$ corresponds to $\theta = 21.79^{\circ}$, $13.17^{\circ}$, $9.43^{\circ}$, and $7.34^{\circ}$, respectively.

Once the matrix representation of equation (\ref{real_bse}) has been built, the third and final step is its numerical diagonalization. This last computation provides us with the excitonic energies and wave functions of the system in consideration. One of the main advantages of the real-space representation is that the excitonic wave functions can be straightforwardly plotted. The following section presents a detailed analysis of the excitonic energies and wave functions.  Note that our method allows us to study systems with around two thousand atoms in a few seconds, which is impossible when employing higher-level theories \cite{biswas2023py}.

\section{Excitonic wave functions of twisted hBN}
\label{excitonic_wf}

This section presents the application of the  abovementioned model  defined for commensurated twisted AA' hBN bilayers. Since excitonic effects play an important role in the optical spectrum of 2D semiconductors \cite{ross2013electrical,xiao2017excitons,mak2018light}, including hBN, the characterization of the excitonic states is an essential ingredient for technological applications. Accordingly, we study the symmetry properties of excitonic wave functions obtained from the effective two-particle Hamiltonian in equation (\ref{real_bse}) for different rotation angles, $\theta$. Before that, it is convenient to understand how the excitonic wave function can be obtained from the matrix representation, in the $\left | \bm{R}\right>$ basis, of equation (\ref{real_bse}). Any eigenvector of this matrix, with eigenvalue number $\lambda$, can be written as $n_h$ blocks of size $M$, which means, $\bm{\Psi}_{\bm{R}}^{\lambda} = \left(\bm{\Phi}^{\lambda}_{h_{1}}, \bm{\Phi}^{\lambda}_{h_{2}}, \hdots, \bm{\Phi}^{\lambda}_{h_{n_h}}\right)$. Each of these blocks, here represented by the numbers $\bm{\Phi}^{\lambda}_{h_{\alpha}} \equiv \left \{\bm{\Phi}^{\lambda}_{\bm{r}_{h_{\alpha}},\bm{r}_{e_1}}, \hdots, \bm{\Phi}^{\lambda}_{\bm{r}_{h_{\alpha}},\bm{r}_{e_M}}\right\}$, stands for all the electron-hole pairs which can be formed with a fixed hole $h_{\alpha}$ located at $\bm{r}_{h_{\alpha}}$ (with $\alpha = 1, \hdots, h_{n_h}$) and an electron located at $\bm{r}_{e_{\beta}}$, with $\beta$ running over all the positions in the system, in other words, $\beta = 1, \hdots, M$. As a result, the total excitonic wave function, equation (\ref{realexc_wf}), is obtained by adding up these $n_h$ parts. Remarkably, on the $\left| \bm{R}\right>$ basis, the excitonic wave function can easily be plotted in real space, enabling the characterization of its symmetry properties.

\begin{figure*}
\includegraphics[scale=0.7]{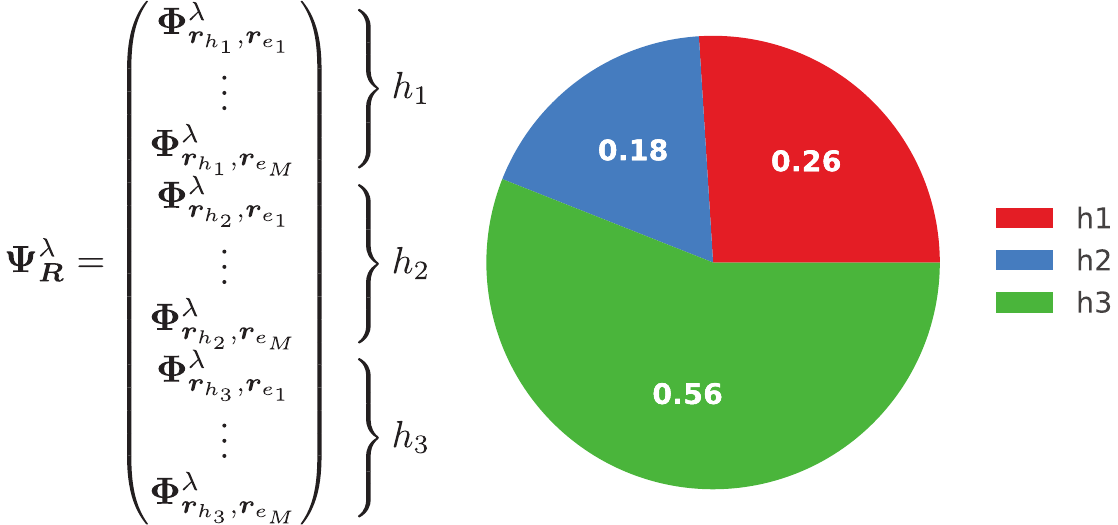}
\caption{Schematic representation of the second ($\lambda = 2$) eigenvector of the effective two-particle Hamiltonian of the BSE obtained for a twisted AA' hBN bilayer with $(m_1,m_2) = (2,1)$. Its mathematical form is shown on the left of the figure. The full circle represents the total excitonic wave function, while each color represents the contribution of specific holes to it.  The three non-equivalent holes for this case are displayed in Fig. \ref{holepos54}, where the atomic environment of the holes is completely different from each other, leading to different contributions to the entire excitonic wavefunction.}
\label{exc_wf_pie21}
\end{figure*}

\begin{figure}
\includegraphics[scale=0.9]{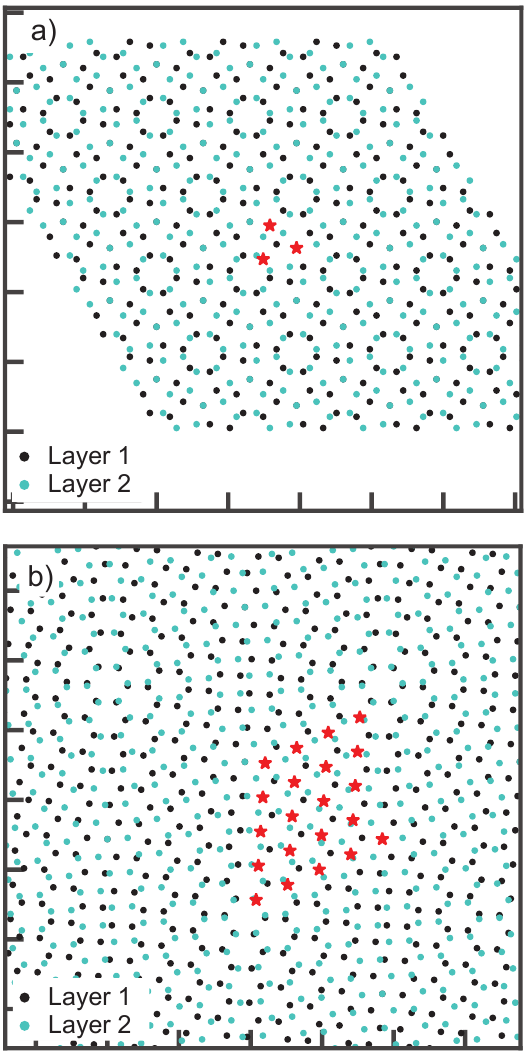}
\caption{Panel a). Non-equivalent position for nitrogen atoms (labeled by red solid stars) of a twisted AA' hBN bilayer with $(m_1,m_2) = (2,1)$. Here and in panel b), atoms belonging to layer 1 (layer 2) are denoted by solid black (solid blue) circles. Panel b). Non-equivalent positions for nitrogen atoms of a twisted AA' hBN bilayer with $(m_1,m_2) = (5,4)$. The non-equivalent positions, twenty-one for this particular case, are labeled by solid red stars.}
\label{holepos54}
\end{figure}

For illustrative purposes, in Fig. \ref{exc_wf_pie21}, a schematic representation of the second ($\lambda = 2$) eigenvector of a twisted AA' hBN bilayer with $(m_1,m_2) = (2,1)$ is displayed. On the left-hand side of the figure, we show a particular case of $\bm{\Psi}_{\bm{R}}^{\lambda}$ for $n_h = 3$ and $\lambda = 2$. As can be seen, that eigenvector is formed by three blocks, each having $M$ items. It is important to remark that even though all the blocks have the same size, they do not equally contribute to the total excitonic wave function. To show this, we plot a pie chart on the right side of Fig. \ref{exc_wf_pie21}, wherein the different colors represent the contributions due to different holes. Each non-equivalent hole contributes differently to the excitonic wavefunction. This latter can be understood by noting that the atomic environment dramatically affects the potential that acts over the exciton. Therefore its contribution to the excitonic wave function will be unique. Note how the complexity of $\bm{\Psi}_{\bm{R}}^{\lambda}$ grows as the number of the non-equivalent position for the holes increases.  In particular, for a twisted AA' bilayer with $(m_1,m_2) = (5,4)$, we have that $n_h = 21$, as seen in Fig. \ref{holepos54}. Even for this seemingly small number of atoms and a supercell of size $3\times3$, the computational resources needed for the diagonalization of the matrix representation of equation (\ref{real_bse}) is considerable. For us, studying angles smaller than $\theta = 6^{\circ}$ is out of reach  because the number of Hamiltonian matrix elements will be $\approx 10^9$ or larger, demanding a massive amount of RAM and CPU time exceeding several months.

\begin{figure}
\includegraphics[scale=1]{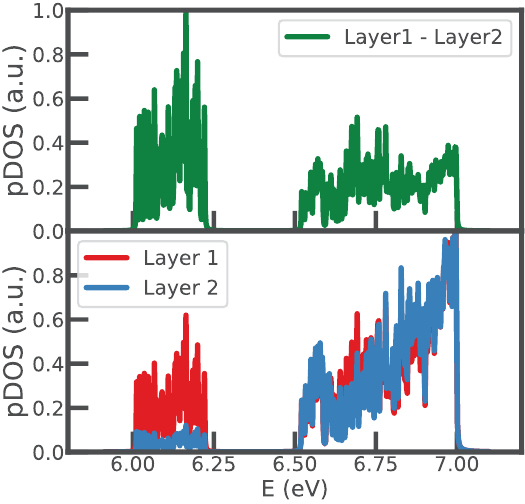}
\caption{Projected  excitonic DOS of a twisted AA' hBN bilayer with $(m_1,m_2) = (5,4)$ is shown. In the lower panel, solid red (blue) lines indicates the DOS projected over the first (second) layer.  In the upper panel,the difference between the projected DOS over the first layer and the second one is presented in solid green line. Intralayer states are characterized by having a very large projection at a single layer. On the other hand, interlayer states have similar projections over the two layers.}
\label{2p_pDOS54}
\end{figure}

\subsection{Excitonic wavefunction for $\theta = 7.34^{\circ}$}

For convenience, we have chosen to put on view only one case, specifically, $(m_1,m_2) = (5,4)$, which corresponds to a rotation angle of $\theta = 7.34^{\circ}$.  We have selected this system because it is the smaller angle we can handle. Besides that, the main features of it are very similar to the ones that appear at other angles. Before showing the excitonic wave functions, it is interesting to look at the intralayer or interlayer character of the excitonic states of this particular case. With this aim, we plotted the projected two-particle DOS, over different layers, of the excitonic energies of a $(5,4)$-twisted hBN bilayer in Fig. \ref{2p_pDOS54}.  It is noteworthy that this density of states was calculated using the excitonic energies weighted by the different contributions of the excitonic wavefunctions in real space considering a distance cutoff of $7$ \AA. We plotted the DOS projected over the first (second) layer in solid red (solid blue) lines in the lower panel of such a figure.

In contrast, the difference between both is exhibited in solid green lines in the upper panel. By inspecting the lower panel of Fig. \ref{2p_pDOS54}, one can elucidate a given excitonic energy of the interlayer or intralayer nature. Excitonic states with a strong intralayer character are almost at a single layer. Thus, the projected DOS must be more significant in one of the two layers. Contrary, interlayer excitonic states are distributed among the two layers of the system. This latter leads to a projected DOS having similar contributions on both layers. For a $(5,4)$ twisted AA' hBN bilayer, the excitonic states with the lowest energies (from $6$ eV to approximately $6.25$ eV) have a strong intralayer character, whereas, for energies greater than $6.25$ eV, they exhibit a robust interlayer nature, see Fig. \ref{2p_pDOS54}. It has to be noted that due to the symmetry reduction imposed by the rotation of the layers, all the excitons, even the ones appearing at the lowest excitonic energies, become non-degenerate. This latter contrasts with the untwisted case where most low-energy and bright excitons doubly degenerate. This behavior is typical, at least, for the $\theta$ angles we have studied.

\begin{figure*}
\includegraphics[scale=0.9]{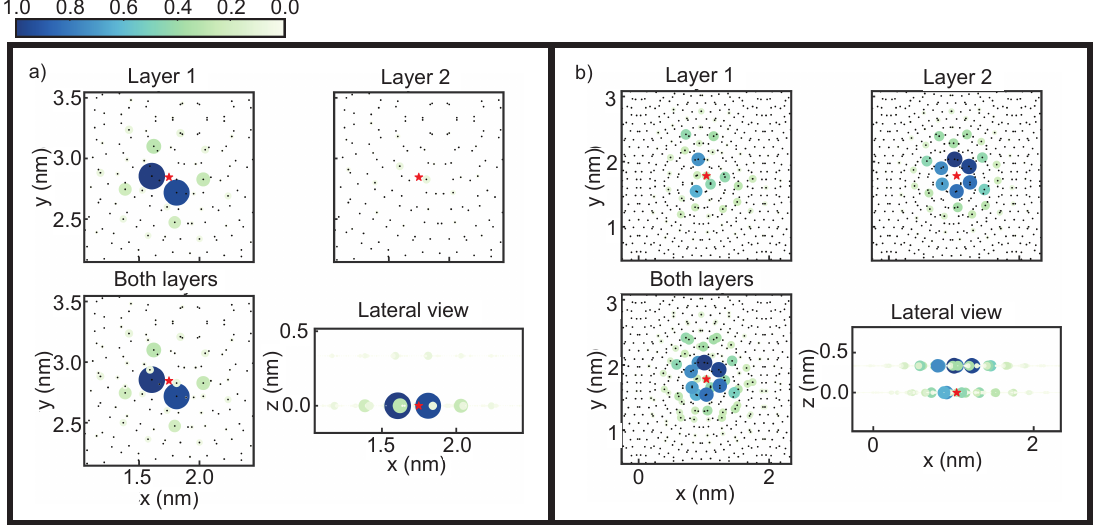}
\caption{Excitonic wave functions of a $(5,4)$-twisted AA' hBN bilayer for two excitonic energies, with eigenvalue numbers $\lambda = 23$ and $\lambda = 55$ for the left, a), and right, b), panels, respectively. Each of these panels shows the excitonic wave function $\Phi_{h_{\beta}}^{\lambda}$. Here we have chosen $\beta = 5$ [for a) panel] and $0$ [for b) panel], projected on layer 1, on layer 2, and on  both layers. Also, a lateral view is included to better appreciate the intralayer or interlayer character of the excitonic state. Solid circles indicate the site of the wave function in real space. Colors represent the square of the wave function amplitude, namely, $\left |\Phi_{h_{\beta}}^{\lambda}\right |^2$ at each atomic position. Finally, the specific position of the hole is indicated by a solid red star.}
\label{wf54}
\end{figure*}

We present two excitonic wave functions obtained for a $(5,4)$-twisted AA' hBN bilayer. For demonstrative purposes, we have chosen an intralayer and an interlayer excitonic state to be analyzed. The $\Phi_{h_{\beta}}^{\lambda}$ part of the excitonic wave functions for $(\beta,\lambda) = (6,24)$ and $(1,56)$ is shown in the corresponding a) and b) panels of Fig. \ref{wf54}. Note that the excitonic wave function of the panel a) is strongly localized at a single layer, the one where the hole is fixed (hence its intralayer nature) and that it resembles the symmetry of a single component of the first exciton appearing in a normal hBN bilayer, which is also of an intralayer nature (see, for example, reference \cite{galvani2016excitons}). As a result, the symmetry of this exciton is reduced for its degenerate version. The exciton presented in panel a) of Fig. \ref{wf54} has the same symmetries as the antisymmetric component of the ground state exciton of a non-twisted AA' hBN bilayer.

On the other hand, the higher energy exciton, shown in panel b) of Fig. \ref{wf54}, has a greater spatial extension than the one appearing in panel a) on the same figure. As easily verified, this exciton has an interlayer character since most of its wave function is located at the layer with no holes.  In addition,  observe that this exciton's structure is more complex than the one of panel a). A more detailed study of the symmetry properties of these excitons is of great interest but is out of this work's scope and will be done elsewhere.

\subsection{Absorption spectrum for $\theta = 7.34^{\circ}$}

In order to know which of these excitonic states is bright or dark, one has to look at the absorption spectrum of the system, given by the imaginary part of the macroscopic dielectric function \cite{rohlfing2000electron}, which in the many-body perturbation case  and within the dipole approximation becomes \cite{cardona2005fundamentals},
 \begin{equation}
     \epsilon(\omega) = \frac{8\pi^2e^2}{V}\sum_{\lambda}\left |\sum_{\bm{R}}\Psi_{\bm{R}}^{\lambda}\,d_{\bm{R}}\right |^2\delta{(\omega - E^{\lambda})}
     \label{absorptioneq}
 \end{equation}
 with the dipole element matrix given by \[d_{\bm{R}} = \frac{i\hbar}{m}\bm{n}\cdot \frac{\left< \bm{R}cv \right | \bm{p} \left | \bm{R}'cv \right >}{E^{\lambda}},\]
where $e$ is the electron charge and $m$ its mass, $V$ is the volume of the unit cell, $E^{\lambda}$ are the excitonic energies, $\Psi_{\bm{R}}^{\lambda}$ are the excitonic weights of the $\lambda$-th eigenvector, $\bm{n}$ is the direction of the electric field, and $\bm{p}$ is the momentum operator.

\begin{figure*}
\includegraphics[scale=1.5]{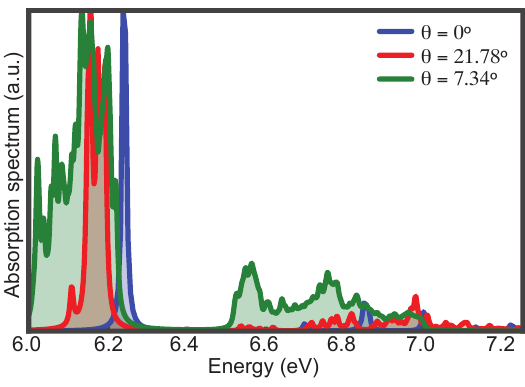}
\caption{Absorption spectra for three different rotation angles, $\theta$, were obtained from the numerical evaluation of equation (\ref{absorptioneq}) using a broadening of $5$ meV. As can be seen, our model qualitatively reproduces the biding energies of the {\it ab initio} calculations. See, for example, Refs. \cite{sponza2018direct,paleari2018excitons}. Note how the rotated cases exhibit a bright exciton inside the band gap of the non-twisted bilayer.}
\label{absortion}
\end{figure*}

The quantity $\epsilon(\omega)$ shows a peak whenever an optical transition is allowed; therefore, the excitons at that energy shall be bright. Figure \ref{absortion} displays the absorption spectra for the three rotation angles studied, $\theta = 0^{\circ}$, $21.78^{\circ}$, and $7.34^{\circ}$. Our model agrees with the binding energies prior obtained from first-principles calculations for the non-twisted hBN bilayer \cite{sponza2018direct,paleari2018excitons}. For this case, $\theta = 0^{\circ}$, the spectrum shows few very well-defined peaks at specific energies, the more intense at about $6.25$ eV, or, equivalently, at an excitonic binding energy of $-1.62$ eV. In particular, this peak is very narrow compared to those that emerged in twisted bilayers. This exciton's characteristics might be attributed to the fact that the unit cell is the smallest, being the more symmetric so that the non-twisted hBN bilayer has more degenerated eigenstates. Therefore, a precise energy source should be used to observe such exciton. It is important to mention that in previous works that used time-dependent DFT, see for example \cite{suzuki2020excitons,camarasa2023transferable}, a shift of the excitonic energies is observed for a non-twisted hBN bilayer when compared to GW-BSE calculations. In \cite{suzuki2020excitons}, the authors claimed that the red-shift of the excitonic energies they have obtained could come from the small vacuum region they used in their calculations or from the fact that they did not use a 2D Coulomb truncation method. Additionally, the authors concluded, among other things, that time-dependent DFT underestimates the excitonic binding energy compared to many-body perturbation theory techniques. In reference \cite{camarasa2023transferable}, a blue-shift of the excitonic energies is observed by using time-dependent DFT altogether with screened range-separated hybrid functionals. This blue-shift of the excitonic peaks, as the authors claimed, it is similar in nature to the one observed in \cite{ramasubramaniam2019transferable}. However important this could be, it is out of the reach of the present work and it is not discussed here. On the other hand, the absorption spectra for bilayers wit $\theta \neq 0^{\circ}$ exhibit more and broader peaks, where the more intense ones are at lower energies than the untwisted counterpart. Notice that for smaller angles, wider excitonic energy windows are observed. Also, as the angle decreases, the unit cell is larger, and the eigenstates are less degenerated, giving rise to an energy dispersion of them. Therefore,  excitonic peaks are observed that become wider as the angle diminishes. For these cases, the excitons can be excited by a range of energies given by the width of the excitonic peak. Third, the effect is even greater for the angle $\theta=7.34^{\circ}$, for which the dominant peaks appear even at lower energies when compared to the $\theta = 0^{\circ}$ and $\theta = 21.78^{\circ}$ cases. This latter could be a consequence of the dispersionless states that appear in the band structure of the twisted hBN bilayer. Due to the low kinetic energy of these states, interaction effects are stronger, enhancing the excitonic properties of the system. Since the moir\'e potential can be strong near the hole, the exciton could be trapped, giving rise to well-localized excitons. The localization of the excitons can be studied readily within this formalism. However, its discussion is out of the scope of this work and will be discuss elsewhere.

Ultimately, let us discuss the applicability and limitations of our model. The main limitations of our model are the approximations that were made, which include the optic limit approximation ($\bm{Q} = 0$), the neglect of quasiparticle and exchange effects, the use of single-particle wave functions that do not have the symmetry of the systems, the use of a model potential instead of using a self-consistent one, and the supposition that holes are only located at nitrogen atoms. Even though this approximation works well for non-twisted hBN bilayers, their validity on twisted systems must be clarified. However, we expect our model to be good for semiconductors with a wide band gap whose holes and electrons tend to be in different sublattices, suppressing the exchange effects. On the other hand, quasiparticle corrections can be included via the single-particle Hamiltonian by fitting it to the GW calculation of the untwisted system as a first approximation. Also, a fully symmetrized set of the wave function can be chosen as the basis on which the one-particle Hamiltonian is to be represented. Notably, our approach can be used in several 2D systems without significant modifications, which makes it useful for systematically studying their optical properties, at least under the above approximations. 

\section{Conclusions}
\label{sec:conclusions}

To summarize, we have introduced a tight-binding model to solve the Bethe-Salpeter equation (BSE) using an effective potential for electron-hole interaction. This approach was used for studying the excitonic properties of 2D twisted bilayers, in particular, and as an example, we have studied the case of twisted AA' hBN bilayers. Our method solves the BSE in real space using a basis of excitonic transitions. To build such a basis, we used a parameterized one-particle $p_z$ tight-binding Hamiltonian, adjusted to correctly reproduce the band gap at the $\bm{\Gamma}$ point obtained from GW calculations. The main advantage of this approach is that the number of electron-hole pairs needed to solve the BSE can be highly diminished by considering the system symmetries. This approach is possible because the number of holes forming the electron-hole pairs can be set to be the number of non-equivalent nitrogen atoms, where we have supposed the hole states are mainly localized. To further simplify the problem, we have approximated the Bethe-Salpeter kernel by a model potential of the 2D Keldysh type, which is a good approximation for 2D semiconductors with a big band gap. 

Thus, we obtain the excitonic properties of twisted hBN bilayers with a rotation angle of down to $\theta = 7.34^{\circ}$, which is far beyond the limits of the {\it ab initio} version of the BSE in reciprocal space. The results obtained from it are expected to be in the range of the experimental error, allowing their qualitative description even the several approximations of our model. Furthermore, the approximations we have introduced can be lifted to improve the model due to its modular structure, albeit at a much higher computational cost.  As we have shown, the excitonic properties of the hBN bilayer can be tuned by controlling the angle between layers, opening the possibility
of novel applications in the far ultraviolet optoelectronic devices. We hope our work motivates more studies in this direction since it allows the study of large moir\'e patterns with relatively moderate computational resources.

Additionally, we have made our implementation of the model in Python code available under the free software license (see the {\it Tight-binding version of the Bethe-Salpeter equation for 2D semiconductors} repository on GitHub.) for anyone who wants to use it. Besides that, we claim our model can be straightforwardly applied to other 2D semiconductors similar to hBN as long as suitable parameterized single-particle tight-binding Hamiltonians and a smooth 2D model potential are
available, thus, opening the possibility of systematically studying their excitonic properties.

\acknowledgements

The authors acknowledge support by the projects PAPIIT IN104622 and IN110023, and by CONACYT Grants No. A1-S-14407 and No. 1564464. P.R.T. also acknowledges the CONACYT for the postdoctoral grant from project No. A1-S-14407. E.O.C. acknowledges financial support from the UNAM via the PRIDIF2021 grant.

\appendix
\section{Details on the fitting tigh-binding parameters} \label{sec:appendix}

\begin{figure*}
\includegraphics[scale=1.0]{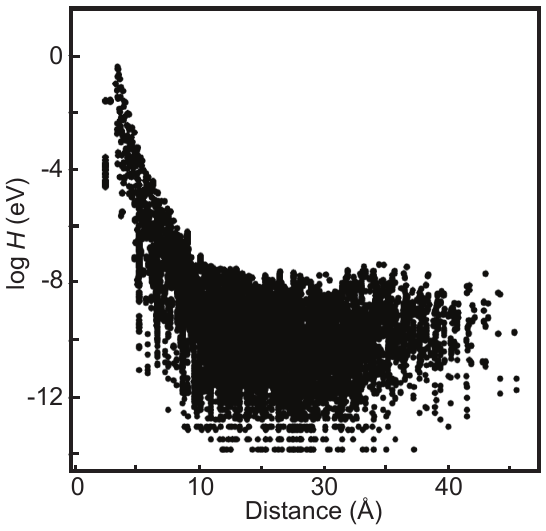}
\caption{Log scale of the matrix elements of the tigh-binding like Hamiltonian, as a function of distance, obtained by the fitting of the Slater-Koster functions to DFT calculations for a (3,2) twisted hBN bilayer. Note that the matrix elements are quite small for distances greater than 7 \AA.}
\label{wan_hoppings}
\end{figure*}

We have used the Pybinding \cite{pybinding} and Pymatgen \cite{pymatgen} packages to construct  the single-particle Hamiltonian. The algorithm used to build the twisted AA' hBN bilayer follows the implementation introduced in the references \cite{aragon2019twisted,sanchez2019unfolding}. For single-particle band structure calculations, we have used $M_{\bm{k}} = 50$ $\bm{k}$-points. We have used a histogram algorithm with $600$ bins and a gaussian envelopment adjusted to reproduce the histogram data best, to obtain the one-particle DOS. As mentioned in the main text, all interlayer interactions were set to zero for distances larger than a radio cutoff of $7$ \AA. To fully justified this assumption a few words are needed. Our tight-binding model was fitted to DFT calculations for both the regular hBN bilayer and the twisted ones. This was made by following the next procedure. First, we have performed DFT calculations for twisted hBN bilayer with indexes (2,1), (3,2), and (4,3). Then, we have extracted a tight-binding Hamiltonian by projecting the plane-wave wavefunctions into maximally localized Wannier functions using Wannier90. We have taken into account only $p_z$ orbitals. Since these orbitals are orthogonal to the rest, the disentanglement procedure was fairly simple. After this, a model was obtained by treating the resulting tight-binding like Hamltonian, particularly by truncating it to 7 \AA$\,$ interactions, and then by fitting the corresponding Slater-Koster functions. For instance, for the (3,2) twisted hBN bilayer system, the resulting maximally localized wannierized Hamiltonian was found to have the matrix elements that are shown in figure \ref{wan_hoppings}. Therein, one can see that the Hamiltonian elements are very low after a few \AA's. For this reason, we have truncated the analysis to 7 \AA.

\begin{figure*}
\includegraphics[scale=1.0]{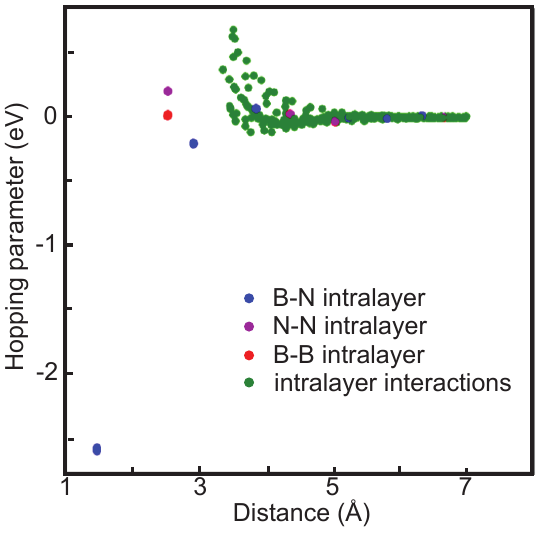}
\caption{Intralayer hoppings for BN interactions (solid blue circles), NN interactions (solid violet circles), and BB interactions (solid red circles). The three types of interlayer hopping parameters (BN, BB, and NN interactions) are displayed as solid green circles, see figure \ref{inter_hoppings}.}
\label{all_hoppings}
\end{figure*}

\begin{figure*}
\includegraphics[scale=1.0]{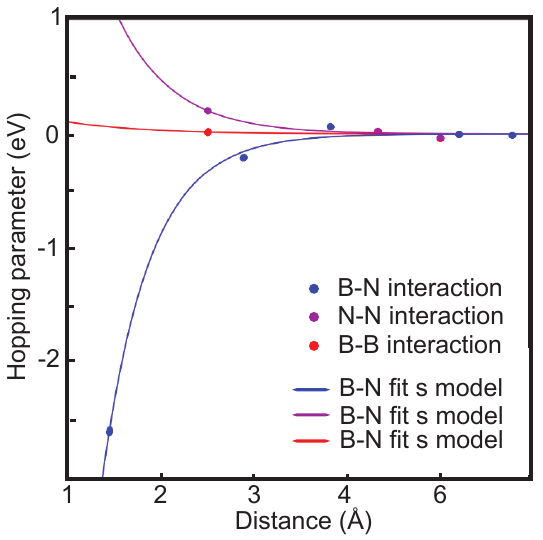}
\caption{Fitting of the intralayer hopping parameters to a model involving only $s$ orbitals. Note that this type of model is in very good agreement with the DFT results.}
\label{intra_fit}
\end{figure*}

\begin{figure*}
\includegraphics[scale=1.0]{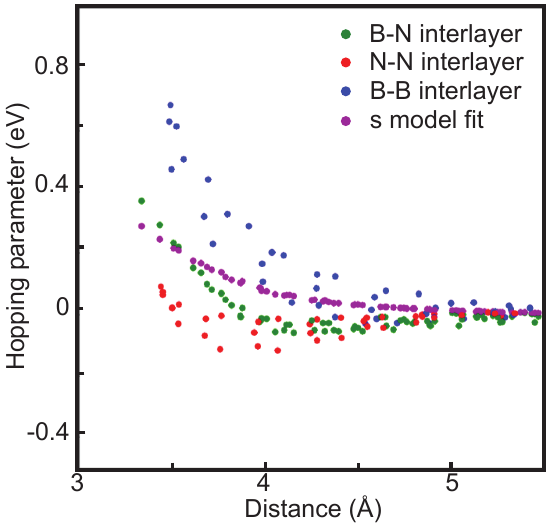}
\caption{Interlayer hopping parameters for BN (solid green circles), NN (solid red circles), and BB (solid blue circles) interactions. As can be seen a model with only $s$ orbitals (solid violet circles) is not enough to fully account for the distance dependence of these hopping parameters, see figures \ref{inter_bnfit}, \ref{inter_bbfit}, and \ref{inter_nnfit}.}
\label{inter_hoppings}
\end{figure*}

\begin{table}[!htbp]
\label{T:equipos}
\begin{center}
\begin{tabular}{| c | c | c | c | c |}
\hline
 \bf{Interactions} & \multicolumn{4}{ c |}{\textbf{Errors}}  \\ 
\cline{2-5}
& $V_{pp\sigma}$ & $V_{pp\pi}$ & $\beta_1$ & $\beta_2$ \\
\hline

Intralayer BB & - &  0.044 & - & - \\ \hline
Intralayer NN & - &  0.155 & - & - \\ \hline
Intralayer BN & - &  0.007 & - & - \\ \hline
Interlayer BB & 0.022 & 0.189 & 0.513 & 0.575 \\ \hline
Interlayer NN & 0.038 & 0.573 & 12.272 & 0.863 \\ \hline
Interlayer BN & 0.040 & 0.055 & 0.253 & 0.124 \\ \hline

\end{tabular}
\caption{Errors of the hopping parameters of the $p_z$ tight-binding model at one standard deviation for a (3,2) twisted hBN bilayer.}
\label{table_par}
\end{center}
\end{table}

\begin{figure*}
\includegraphics[scale=1.0]{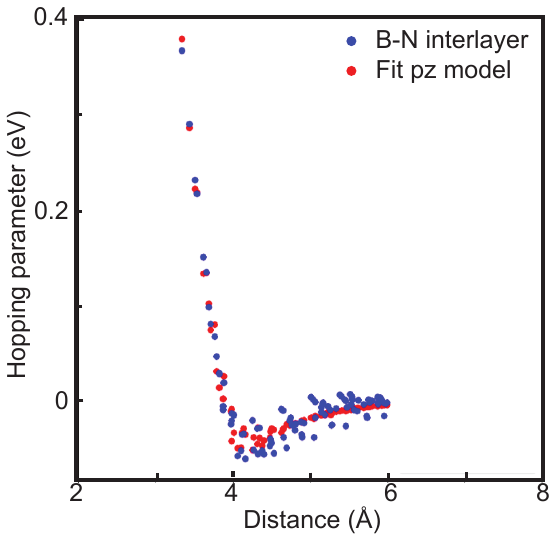}
\caption{Fitting of the interlayer BN hopping parameters (solid blue circles) to a $p_z$ model (red solid circles).}
\label{inter_bnfit}
\end{figure*}

\begin{figure*}
\includegraphics[scale=1.0]{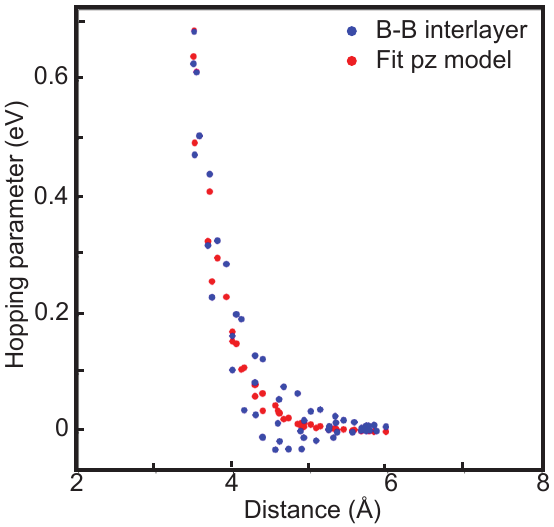}
\caption{Fitting of the interlayer BB hopping parameters (solid blue circles) to a $p_z$ model (red solid circles).}
\label{inter_bbfit}
\end{figure*}

\begin{figure*}
\includegraphics[scale=1.0]{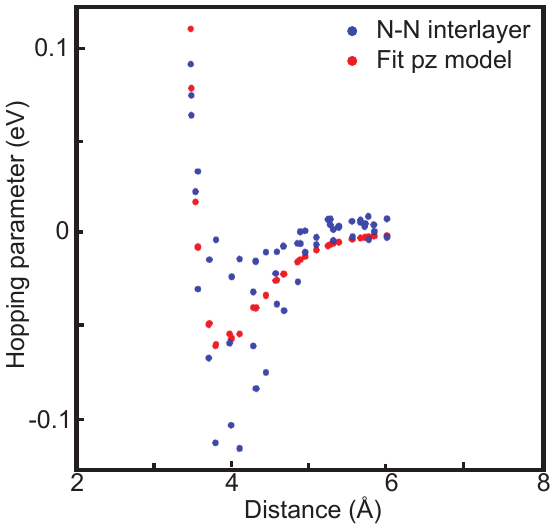}
\caption{Fitting of the interlayer NN hopping parameters (solid blue circles) to a $p_z$ model (red solid circles).}
\label{inter_nnfit}
\end{figure*}

Since the Wannier centers are localized fairly close to the atomic positions a simple script can be used to separate the different contributions of the various types of interactions, typical results can be seen in figure \ref{all_hoppings}. We have approximated all the intraleyer hopping parameters by using a $s$-type model, as can be seen in figure \ref{intra_fit} this type of model correctly reproduces the DFT calculations. On the other hand, interlayer hopping parameters are more complex. From figure \ref{inter_hoppings}, it can be clearly seen that a model using only $s$ orbitals does not describe correctly their distance dependence (see the solid violet circles in that figure), as a matter of fact, a $p_z$ model is needed to rightly reproduce the DFT results. The corresponding fitting of this kind of hopping parameters is shown in figures \ref{inter_bnfit}, \ref{inter_bbfit}, and \ref{inter_nnfit} for BN, BB, and NN interactions, respectively. Although the agreement between DFT results and the $p_z$-type tight-binding Hamiltonian is quite good, a certain amount of variations are observed in the previous figures. To quantify the goodness of such a fitting we can compute the errors up to one standard deviation for each twisted hBN bilayer. The resulting error for the three types of interlayer interactions, for a (3,2) twisted hBN bilayer, are summarized in table \ref{table_par}.

\begin{figure*}
\includegraphics[scale=1.0]{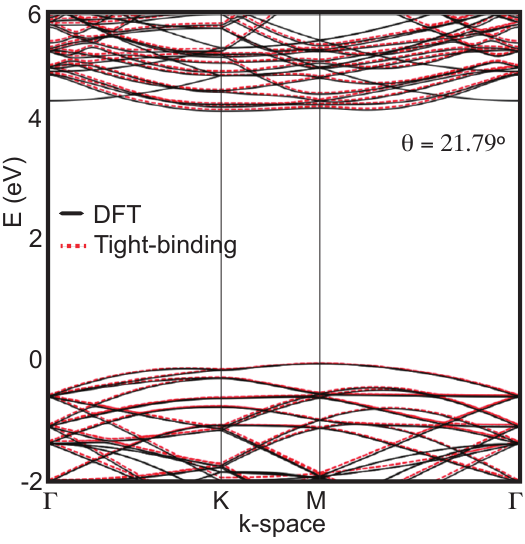}
\caption{Comparison of the band structures of a (3,2) twisted hBN bilayer obtained from DFT calculations (solid black lines) and from the $p_z$-type tight-binding model (dotted red lines). Note the excellent agreement between them.}
\label{bands_comp}
\end{figure*}

It is time to compare the band structures of a (3,2) twisted hBN bilayer for both the DFT calculations and the $p_z$-type tight-binding model. This is done in figure \ref{bands_comp}. Therein, it can be seen that the $p_z$ model is in very good agreement with the DFT band structure thus proving that this type of model is adequate to study rotated systems.

Concerning two-particle calculations, a $5\times5$ supercell was used to compute the excitonic properties of twisted AA' hBN bilayer with $(m_1,m_2) = (2,1)$ and $(3,2)$, whereas a $3\times 3$ supercell was used for twisted systems having $(m_1,m_2) = (4,3)$ and $(5,4)$. The number of atoms for $(m_1,m_2) = (2,1)$, $(3,2)$, $(4,3)$, and $(5,4)$ are $M = 700$, $1900$, $1332$, and $2196$, respectively. We have used the Spglib library \cite{togo2018texttt} via Pymatgen to compute the non-equivalent positions of the twisted systems studied here. For the 2D Keldysh potential, we have used a radio cutoff ($r_{c}$) of twice the size of the unit cell of the twisted system. To be specific, $r_c = 1.3$ nm, $2.1$ nm, $3.0$ nm, and $3.8$ nm for rotation angles of $\theta = 21.79^{\circ}$, $13.17^{\circ}$, $9.43^{\circ}$, and $7.34^{\circ}$. Finally, the excitonic energies and  weights of the two-particle wave functions were obtained by direct diagonalization of the effective two-particle Hamiltonian of the equation (\ref{real_bse}), which was built using our algorithm.

\bibliography{bibliography}{}

\end{document}